\documentclass[preprint,aps,12pt,preprintnumbers,eqsecnum,nofootinbib]{revtex4}
\usepackage{graphicx}
\usepackage{subfigure}
\usepackage{amssymb,amsmath,amsfonts}
\usepackage[english]{babel}

\usepackage{color}
\usepackage{amssymb,amsmath}

\newcommand{\beq}{\begin{equation}}
\newcommand{\enq}{\end{equation}}
\newcommand{\ones}{\mathbf{1}_S}
\newcommand{\onea}{\mathbf{1}_A}
\newcommand{\two}{\mathbf{2}}
\newcommand{\phiang}{\left(\frac{3\theta}{v}\right)}

\begin{document}
% 
% version revised for the referee
%
% Title of paper
\title{\vspace*{0.5in} 
A Permutation on Hybrid Natural Inflation
\vskip 0.1in}
\author{Christopher D. Carone}\email[]{cdcaro@wm.edu}
\author{Joshua Erlich}\email[]{jxerli@wm.edu}
\author{Raymundo Ramos}\email[]{raramos@email.wm.edu}
\author{Marc Sher}\email[]{mtsher@wm.edu}
\affiliation{High Energy Theory Group, Department of Physics,
College of William and Mary, Williamsburg, VA 23187-8795}
%=
\date{July 22, 2014}

\begin{abstract}
We analyze a model of hybrid natural inflation based on the smallest non-Abelian discrete group $S_3$.   Leading invariant terms in the
scalar potential have an accidental global symmetry that is spontaneously broken, providing a pseudo-Goldstone boson that is identified as 
the inflaton.  The $S_3$ symmetry restricts both the form of the inflaton potential and the couplings of the inflaton field to the waterfall fields 
responsible for the end of inflation.  We identify viable points in the model parameter space.  Although the power in tensor modes is  small in 
most of the parameter space of the model, we identify parameter choices that yield potentially observable values of $r$ without super-Planckian
initial values of the inflaton field.
\end{abstract}
\maketitle

\section{Introduction} \label{sec:intro}

Measurements of the anisotropy in the cosmic microwave background (CMB) have led to the development of a ``standard model" of cosmology, 
with a cosmological constant, cold dark matter and a spectrum of initial CMB fluctuations that seed large scale structure~\cite{Dodelson:2003ft}.   It is widely 
believed that these initial fluctuations arise from an inflationary epoch, resulting in a nearly scale-invariant spectrum.    More precise measurements of the CMB 
fluctuations, including polarization measurements, have been carried out by experiments such as WMAP~\cite{Hinshaw:2012aka}, PLANCK~\cite{Ade:2013uln} and 
BICEP2~\cite{Ade:2014xna}.  These measurements provide information about initial metric perturbations that can severely constrain (or rule out) various 
inflationary models.

In order to satisfy the limits on the size of the CMB anisotropy fluctuations, the scalar self-coupling constant of the inflaton field must be very small, 
typically less than $10^{-12}$ in most realistic models~\cite{Adams:1992bn}.    While such a small coupling could be assumed, it would be 
aesthetically more desirable if it arose naturally.   This is the case for theories in which the inflaton is identified with the pseudo-Goldstone boson of a 
spontaneously broken approximate global symmetry.  Such ``natural inflation" scenarios were proposed first by Freese, Frieman and 
Olinto~\cite{Freese:1990rb}.  If the scale of spontaneous symmetry breaking is $f$ and if there is an explicit breaking of the global symmetry
via an anomaly, the inflaton potential takes the form
\begin{equation}
V = V_0 \left[ 1 \pm \cos(n\phi/f)\right] \,\,\, ,
\end{equation}
where $n$ is an integer. The model is consistent with measured values of the spectral 
index and its running, as well as constraints on the ratio of powers in tensor and scalar modes~\cite{Freese:2008if}.     A concern about natural inflation 
is that the value of $f$ must be very close to or above the Planck scale, so that quantum-gravitational corrections to the potential are not automatically 
under control.

A model that can result in a lower value of $f$ is ``hybrid natural inflation"~\cite{Stewart:2000pa,Cohn:2000hc,Ross:2009hg,Ross:2010fg}.   The original 
hybrid inflation model, proposed by Linde~\cite{Linde:1993cn}, has a second scalar field which couples to the inflaton and ends the inflationary epoch.   As the 
inflaton slowly rolls, the parameters of the potential of the second scalar field change due to the coupling, and at some point the second scalar field acquires 
a vacuum expectation value, ending inflation.    This second scalar field was referred to as the ``waterfall" field.  Hybrid natural inflation
models are natural inflation models in which inflation is terminated due to the dynamics of such additional fields.

An important question in any model based on the natural inflation idea is the origin of the approximate global symmetry.  Global symmetries
are not believed to be fundamental (for example, they are typically violated by quantum gravitational effects~\cite{Kallosh:1995hi}), so it is desirable to arrange
that these symmetries arise by accident, as a consequence of the form of the leading terms in the potential; these terms  are restricted by the 
continuous or discrete gauge symmetries of the theory.  While discrete gauge symmetries can be thought of as discrete remnants of a
spontaneously broken continuous gauge symmetry~\cite{ibanross}, they also can be defined consistently without such an 
embedding~\cite{bd}; in either case, they are preserved by quantum gravitational effects.  Cohn and Stewart~\cite{Stewart:2000pa,Cohn:2000hc} 
showed that accidental global symmetries could easily be obtained in models with non-Abelian discrete gauge symmetries, and illustrated 
their point with hybrid models based on the discrete group $\Delta(96)$.  They note that many other models based on smaller discrete groups 
are likely possible.  Nevertheless, the literature on such models is relatively sparse.  Ross and Germ\'an~\cite{Ross:2009hg,Ross:2010fg} have 
explored hybrid natural inflation models based on the discrete group $D_4$.   In their model, the inflaton potential takes the form
\begin{equation}
V = V_0 \left[1 + a \cos(\phi/f)\right] \,\, ,
\label{eq:modv}
\end{equation}
where $a$ is a constant.  This potential can generate phenomenologically acceptable inflation with $f$ substantially smaller then the Planck 
mass, so that higher-order corrections are under control.  Ross and Germ\'an~\cite{Ross:2009hg,Ross:2010fg} point out that potentials of the
form Eq.~(\ref{eq:modv}) should be expected in similar models based on other non-Abelian discrete groups.

Given the promise of the models considered in Refs.~\cite{Stewart:2000pa,Cohn:2000hc,Ross:2009hg,Ross:2010fg}, and
motivated by minimality, we explore in this paper a hybrid natural inflation model based on the smallest non-Abelian discrete group,
the permutation group $S_3$.   The discrete symmetry restricts both the inflaton potential and the couplings of the inflaton to the waterfall fields.  
The $S_3$ charge assignments in our model satisfy the requirements for a discrete gauge symmetry, as set out in Ref.~\cite{bd}. 
In Sec.~\ref{sec:s3}, we review the group $S_3$ and its representations.  The model is presented in Sec.~\ref{sec:model}.   After
reviewing inflationary parameters in Sec~\ref{sec:params}, we study a typical point in model parameter space in quantitative detail in Sec.~\ref{sec:num}.   
Motivated by the potential signature in gravitational waves, we show in Sec.~\ref{sec:gw} that the model can yield a potentially observable 
tensor-to-scalar ratio, without requiring super-Planckian values of the inflaton field, and we explain why this is not in conflict 
with the Lyth bound~\cite{Lyth:1996im}.   In Sec.~\ref{sec:offreheat}, we discuss the cutoff of inflation and reheating.  Finally, in Sec.~\ref{sec:conc}, we 
present our conclusions.

\section{The group $S_3$} \label{sec:s3}

We base our model on $S_3$, the smallest non-Abelian discrete symmetry group.   The group has six elements whose action can be identified with the permutation of
three objects.  A useful discussion of this symmetry in a model building context can be found in Ref.~\cite{s3papers}.

$S_3$ has three irreducible representations: a two-dimensional representation $\two$ and two one-dimensional representations, $\onea$ and $\ones$.  The $\ones$
representation is the trivial singlet.  The rules for group multiplication are given by $\onea \otimes \onea = \ones \otimes \ones=\ones$,  $\onea \otimes \ones = \onea$ and
$\two\otimes\two=\two \oplus \onea \oplus \ones$.   The product of two doublet representations can be decomposed into its irreducible components using Clebsch-Gordan
matrices.  Let $\psi$ and $\eta$ represent two-component column vectors that transform as doublets under $S_3$ and let $\sigma^a$ denote the Pauli matrices.   
The products $\psi^T C_{\ones} \eta$ and $\psi^T C_{\onea} \eta$ transform in the $\ones$ and $\onea$ representations, respectively, where
\begin{equation}
C_{\ones} = \openone \,\,\,\,\, \mbox{       and       } \,\,\,\,\, C_{\onea} = i \sigma^2 \,\,\, .
\end{equation}
Similarly, we can construct a doublet 
\begin{equation}
\left[ \begin{array}{c} 
\psi^T C^{(1)}_{\two} \eta \\  \psi^T C^{(2)}_{\two} \eta 
\end{array} \right]  \sim \two \,\,\, ,
\end{equation}
where
\begin{equation}
C^{(1)}_{\two} = \sigma^3 \,\,\,\,\, \mbox{       and       } \,\,\,\,\, C^{(2)}_{\two} = - \sigma^1 \,\,\, .
\end{equation}

The model we present in the next section includes an $S_3$ doublet field $\phi=(\phi_1,\phi_2)^T$, so it is useful to enumerate the $S_3$ invariants
that can be constructed from products of $\phi$, up to quartic order.  The quadratic combination of fields that transforms in the $\ones$ representation has the form
\begin{equation}
(\phi^2)_{\ones} \equiv \phi^T C_{\ones} \phi = \phi_1^2 + \phi_2^2 \,\,\, .
\label{eq:ppin}
\end{equation}
While there are three $\ones$ reps in the product $\two \otimes \two \otimes \two \otimes \two$, all such invariants constructed from a single $\phi$
have the same form,
\begin{equation}
(\phi^4)_{\ones} = (\phi_1^2 + \phi_2^2)^2 \,\,\,.
\label{eq:ppppin}
\end{equation}
While Eqs.~(\ref{eq:ppin}) and (\ref{eq:ppppin}) follow from $S_3$ invariance, it is important to note that
these expressions are also invariant under a continuous symmetry, SO(2), under which the $\phi$ field is also
a doublet.  However, this accidental symmetry is broken by the $S_3$ cubic invariant
\begin{equation}
(\phi^3)_{\ones}  = \phi_1 \, (\phi_1^2 - 3 \,\phi_2^2)
\label{eq:cubin}
\end{equation}
The model of the next section will identify the inflaton field $\theta$ with the pseudo-Goldstone boson of this accidental SO(2) symmetry;
the soft breaking of this symmetry by the cubic invariant will be used to generate the inflaton potential.   Notice, if we parameterize
\begin{equation}
\phi = (\rho + v) \left[\begin{array}{c} \cos(\theta/v) \\ \sin(\theta/v) \end{array} \right] \,\,\, ,
\label{eq:phiparam}
\end{equation}
where $v$ is the scale of spontaneous symmetry breaking and $\rho$ is the massive radial excitation, then Eqs.~(\ref{eq:ppin})
and (\ref{eq:ppppin}) are independent of $\theta$, indicating that these terms contribute nothing to the inflaton potential. 
(Note that in this parameterization the kinetic term for $\theta$ is canonically normalized.) On the
other hand, Eq.~(\ref{eq:cubin}) simplifies to
\begin{equation}
(\phi^3)_{\ones} = (\rho+v)^3 \cos(3 \, \theta /v) \,\,\,,
\end{equation}
which can be used to lift the flat direction.  In the next section we show how these ingredients can be combined to produce a
viable model of hybrid natural inflation.

\section{The Model} \label{sec:model}

In addition to the doublet field $\phi$ described in the previous section, our model includes two real scalars, $\chi_1$ and $\chi_2$, each in the 
$\ones$ representation of $S_3$.  We assume a $\mathbb{Z}_2$ symmetry under which both $\chi$ fields are odd, which eliminates unwanted
linear terms that would otherwise give the $\chi_i$ vevs.   The SO(2) invariant terms in the potential 
\begin{equation}
V_{\rm SO(2)} (\phi,\chi_i) = -\frac{1}{2} m_\phi^2 (\phi_1^2+\phi_2^2) + \lambda_\phi \, (\phi_1^2+\phi_2^2)^2 + \cdots
\end{equation}
lead to the spontaneous breaking of the SO(2) symmetry due to the negative mass squared term for $\phi$.  The terms not shown include 
various $\phi^2 \chi^2$ couplings as well as the potential for the $\chi_i$ fields by themselves.  It is not hard to see that it is possible to choose 
parameters such that $\phi^2$ develops a vacuum expectation value, while the $\chi_i$ do not.  The details are not crucial for our purposes because 
the SO(2) invariant terms have no effect on the form of the inflaton potential.  All that is relevant at this stage is that the spontaneous symmetry 
breaking is consistent with the parameterization in Eq.~(\ref{eq:phiparam}), with the Goldstone boson $\theta$ 
identified as the inflaton.

In the spirit of a perturbative expansion, we now introduce smaller terms which violate the accidental SO(2) symmetry.  At the renormalizable level, 
we can include a term of the form $m_0 \,(\phi^3)_{\ones}$;  the dimensionful coefficient $m_0$ parameterizes the breaking of the SO(2) symmetry.   We could
simply assume a small value of $m_0$ as a fine-tuning in the model (after all, we have to accept the same for the Higgs boson mass in any
non-supersymmetric theory).  However, we can do better if we allow an additional $\mathbb{Z}'_2$ symmetry under which the $\phi$ doublet and $\chi_1$ 
are odd, and treat $m_0$ consistently as a soft $\mathbb{Z}'_2$-breaking parameter.   Since the $\mathbb{Z}'_2$ 
symmetry is restored in the limit of vanishing $m_0$,  there can be no large radiative corrections and a small $m_0$ will be natural following the 
criterion of t'Hooft~\cite{'tHooft:1979bh}.  We will adopt this assumption henceforth.  The only other term that we include that violates the SO(2) symmetry 
is of the form $\chi_1\chi_2 (\phi^3)_{\ones}$.  Identifying the $\chi$ fields as the waterfall fields of a hybrid inflation model, such couplings are responsible for ending 
inflation in the model.  In the present case, this SO(2) breaking term is Planck suppressed for sub-Planckian field values.

We now consider the effective theory below the SO(2)-breaking scale (the scale of the $\rho$ mass).  With the particle content and the
symmetries of the theory as we have specified them, the scalar potential for the $\theta$, $\chi_1$ and $\chi_2$ fields is somewhat 
cumbersome for a general analysis.  We will therefore adopt a simplifying assumption in our parameter choices to demonstrate  
most simply that viable cosmological solutions exist.  Additional solutions are possible for less restrictive choices of model parameters. 

We study the following simplified form for the scalar potential:
\begin{eqnarray}
V(\theta,\chi_i) &=& V_0 + c_1 \frac{v^3}{M_P} \chi_1 \chi_2 \cos(3 \theta/v) - m_0 v^3 \cos(3 \theta/v) \nonumber \\
&+& \frac{1}{2} m_\chi^2 (\chi_1^2 + \chi_2^2) + (\lambda \chi_1^4 + \lambda_{12} \chi_1^2 \chi_2^2 +\lambda \chi_2^4) \,\, .
\label{eq:simpot}
\end{eqnarray}
Here $V_0$ is a constant, $c_1$, $\lambda$ and $\lambda_{12}$ are couplings, and $m_\chi$ is a common $\chi_i$ field mass. 
The second and third terms are SO(2)-breaking interactions discussed previously.  For definiteness, we assume $c_1>0$.  In contrast 
to the most general case, we have assumed symmetry under $\chi_1 \leftrightarrow \chi_2$.  This simplifying assumption has no effect 
on the shape of the inflaton potential (which is obtained by setting $\chi_i=0$), but substantially streamlines our presentation.
If one relaxes this assumption, one has to contend with minimization conditions that are cubic; this complicates the analysis 
but does not affect our conclusions qualitatively.  Note also that we have omitted the  $(\chi_1^2+\chi_2^2)\cos(3 \theta/v)$ and 
$\chi_1 \chi_2^3+\chi_2 \chi_1^3$ interactions, which are  $\mathbb{Z}'_2$ odd.   Since the $\mathbb{Z}'_2$ symmetry is broken only by $m_0$, these 
are suppressed by $m_0/M_P$ relative to the second and fifth terms in Eq.~(\ref{eq:simpot}), respectively, making them negligible\footnote{If one prefers 
to dispense with the softly-broken $\mathbb{Z}'_2$ symmetry and allow $m_0$ to be fine-tuned, then these terms can be omitted as a parametric simplification.   The effect of 
including a $c_2 \frac{v^3}{M_P}(\chi_1^2 +\chi_2^2)\cos(3\theta/v)$ term, with $c_2>0$, is to change Eq.~(\ref{eq:ends}) by replacing $c_1 \rightarrow c_1 - c_2$ 
and Eqs.~(\ref{eq:chi2}) and (\ref{eq:c0}) by $m_\chi^2 \rightarrow m_\chi^2 + 2 c_2 \frac{v^3}{M_P}$.   If one adds a $\lambda_3 (\chi^3_1 \chi_2 + \chi_2^3 \chi_1)$ 
term, then the only change in these equations is $\lambda_{12} \rightarrow \lambda_{12}-2 \lambda_3$.  These changes do not affect our results qualitatively.}.  We 
set the cosmological constant to zero at the global minimum of the potential by choice of the parameter $V_0$.   

Inflation occurs as the field $\theta$ slow rolls toward the origin, between initial and final field values that lie within the interval $0< 3 \theta/v < \pi$.  
During inflation, the effective $\chi_i$ masses are positive and the $\chi$ fields remain at the origin.   Inflation ends via the waterfall mechanism when $\theta$ is such that
\begin{equation}
c_1 \frac{v^3}{M_P} \cos(3 \theta /v) >  m_\chi^2 \,\,\, .
\label{eq:ends}
\end{equation}
At this point, the $\chi_i$ potential is destabilized and the $\chi$ fields develop vevs\footnote{As we will see in Sec.~\ref{sec:num},
$\cos(3 \theta /v)>0 $ when inflation ends, as has been assumed in Eq.~(\ref{eq:ends}).}.  Within a Hubble time, the fields reach
a global minimum, and inflation abruptly ends.  Oscillations of the waterfall fields about this minimum leads to reheating. 
Given the inequality in Eq.~(\ref{eq:ends}), we find that the location of the degenerate global minima in our model are given by
\begin{equation}
\theta = 0 \,\,\, ,
\end{equation}
\begin{equation}
\chi_1 = - \chi_2  \,\,\, , 
\end{equation}
and
\begin{equation}
\chi_1^2 = \frac{1}{2 \, (2\,\lambda + \lambda_{12})} \left[ c_1 \frac{v^3}{M_P} - m_\chi^2 \right]  \,\,\, .
\label{eq:chi2}
\end{equation}
Setting the cosmological constant to zero at any of these minima determines the constant $V_0$ in Eq.~(\ref{eq:simpot}):
\begin{equation}
V_0 =m_0 v^3 + \frac{1}{4} \frac{1}{(2 \, \lambda + \lambda_{12})} \left( c_1 \frac{v^3}{M_P} - m_\chi^2 \right)^2 \,\,\, .
\label{eq:c0}
\end{equation}
With this result in hand, the form of the inflaton potential during the period of slow roll is fixed in term of the model parameters:
\begin{equation}
V(\theta) = V_0 \left[1-\xi \cos(3 \theta/v) \right]
\label{eq:genpot}
\end{equation}
where $\xi \equiv m_0 v^3 / V_0$ and $V_0$ is given by Eq.~(\ref{eq:c0}).  Our parameter choices in the next sections have $\xi<1$.

Eq.~(\ref{eq:genpot}) is amenable to the standard analysis of a single-field inflation model until the end of inflation.  We review the quantities of interest in such an analysis in the
next section and explore numerical results for a number of benchmark points in our model's parameter space.  For these points, we will also present estimates 
to justify that the shut-off of inflation via the waterfall mechanism is sufficiently fast.

\section{Inflation parameters}\label{sec:params}
In terms of the inflaton potential $V(\theta)$, the slow-roll parameters may be written~\cite{Ade:2013uln}
\begin{equation}
\epsilon\equiv\frac{M_P^2}{16\pi}\left(\frac{V'}{V}\right)^2 \,\, , \,\,\,\,\,\,\,\,\,\,
\eta \equiv \frac{M_P^2}{8\pi}\frac{V''}{V} \,\,\,\,\,\,\,\,\, \mbox{ and } \,\,\,\,\,\,\,
\gamma \equiv \frac{M_P^4}{64 \pi^2} \frac{V^\prime V^{\prime\prime\prime}}{V^2}  \,\, .
\label{eq:srp}
\end{equation}
In a generic single-field model, $\epsilon=1$ is usually chosen to define the end of inflation; in the present case, $\epsilon$ remains small 
throughout the period of slow roll until inflation is terminated by the destabilization of the effective $\chi$ potential.  The number of $e$-folds
of inflation $N$ may be expressed as ~\cite{Dodelson:2003ft}
\begin{equation}
N = \frac{2 \sqrt{\pi}}{M_P} \int_{\theta_f}^{\theta_i} \frac{1}{\sqrt{\epsilon}} \, d\theta  \,\,\, ,
\label{eq:numef}
\end{equation}
where $\theta_i$ and $\theta_f$ are the initial and final inflaton field values, respectively.  We will evaluate this quantity in our model to assure that
sufficient inflation is achieved.

A number of cosmic microwave background parameters can be expressed conveniently in terms of the slow roll parameters, as we now 
summarize~\cite{Ade:2013uln,Beringer:1900zz}.  All are evaluated at values of the inflaton field corresponding to $\sim60$ $e$-folds before the end of inflation, 
when scales of order the current Hubble radius exited the horizon. The amplitude of the tensor power spectrum in the slow-roll approximation is
\begin{equation}
\Delta^2_T(k)=\frac{128}{3}\frac{V}{M_P^4} \,\,\, , 
\label{eq:pt}
\end{equation}
while the amplitude of the scalar power spectrum is 
\begin{equation}
\Delta^2_R(k)=\frac{128\pi}{3M_P^6}\frac{V^3}{V^{\prime 2}}= \frac{8}{3 M_P^4}\frac{V}{\epsilon} \,\,\,. 
\label{eq:ps}
\end{equation}
The ratio of the tensor to scalar amplitudes is then
\begin{equation}
r=16 \, \epsilon. 
\label{eq:r}
\end{equation}  
The scalar spectral index and its running are given by
\begin{equation}
n_s(k) = 1-6\epsilon+2\eta  \,\,\,\,\,  \mbox{ and } \,\,\,\,\, n_r  = 16\,\epsilon\,\eta - 24\epsilon^2-2\gamma \,\, .
\label{eq:ns1}
\end{equation}
The predictions following from our model for the parameters summarized in this section can easily be computed starting with
Eqs.~(\ref{eq:c0}) and (\ref{eq:genpot}).   For example, the slow roll parameters take the form:
\begin{equation}
\epsilon  =\frac{9 \xi ^2 M_P^2 \sin ^2\phiang}{16 \pi  v^2 \left(1-\xi  \cos \phiang \right)^2} \,\,\, ,
\end{equation}
\begin{equation}
\eta  =\frac{9 \xi  M_P^2 \cos \phiang }{8
   \pi  v^2 \left(1-\xi  \cos \phiang \right)} \,\,\, ,
\end{equation}
\begin{equation}
\gamma  = -\frac{81 \xi ^2 M_P^4 \sin ^2\left(\frac{3 \theta
   }{v}\right)}{64 \pi ^2 v^4 \left(1-\xi  \cos \left(\frac{3
   \theta }{v}\right)\right)^2} \,\,\, .
\end{equation}
The parameters $n_s$, $n_r$, $r$ and $\Delta_R^2$ can then be evaluated using these expressions, with $\theta$ set to $\theta_i$ as determined from
Eq.~(\ref{eq:numef}) with $N=60$.   We will follow this procedure in our quantitative analysis in the following section.  The measured values of the 
cosmological parameters that we use in this analysis are $n_s=0.9603 \pm 0.0073$, $n_r = -0.013 \pm 0.009$, 
$r<0.12 \,\,\,(95\%\mbox{ C.L.})$ and $\Delta_R^2 = 2.2 \times 10^{-9}$~\cite{Ade:2013uln}.  Note that the recent observation by the 
BICEP2 experiment of B-mode polarization in the CMB, would imply $r=0.20^{+0.07}_{-0.05}$ if the signal is interpreted as cosmological in origin~\cite{Ade:2014xna}.  
However, the contribution of foreground dust to the BICEP2 signal is currently uncertain, so one cannot draw a reliable conclusion on the value of $r$ from 
this measurement at present~\cite{Mortonson:2014bja,Flauger:2014qra}.  

\section{Numerical Analysis} \label{sec:num}

In this section, we present the numerical analysis corresponding to a typical, benchmark point in the model parameter space.  We will 
find in this example that the primordial gravitational wave signal is small.  In the next section, we show that for a careful choice of 
parameters, a larger value of $r$ can be obtained.

Working with the generic potential, Eq.~(\ref{eq:genpot}), let us focus first on two quantities:  the spectral index,
\begin{equation}
n_s-1 = -\frac{9}{16 \pi}\frac{M_p^2}{v^2} \left\{ \frac{2 \xi^2 \,[2+\sin^2(3 \theta_i/v)]- 4\xi \cos(3 \theta_i/v)}{[1-\xi \cos(3\theta_i/v)]^2} \right\}  \,\, ,
\label{eq:ns}
\end{equation}
and the amplitude of the scalar power spectrum,
\begin{equation}
\Delta_R^2 = \frac{128 \pi}{27} \frac{V_0 v^2}{M_p^6}\frac{1}{\xi^2} \frac{[1-\xi\cos(3\theta_i/v)]^3}{\sin^2(3\theta_i/v)}  \,\, .
\label{eq:dsq}
\end{equation}
Both are evaluated at the initial field value $\theta_i$, corresponding to $60$ $e$-folds before the end of inflation.  The number of $e$-folds, 
following from Eq.~(\ref{eq:numef}),  is given by
\begin{equation}
N = \frac{8 \pi}{9} \frac{v^2}{M_P^2} \frac{1}{\xi} \left[(1-\xi) \ln\left(\frac{\sin(3\theta_i/v)}{\sin(3\theta_f/v)} \right) -
\ln\left(\frac{1+\cos(3\theta_i/v)}{1+\cos(3\theta_f/v)} \right) \right] \,\,\, .
\label{eq:efold}
\end{equation}
Let us define $x_{i,f} \equiv \cos(3 \theta_{i,f}/v)$, as well as
\begin{equation}
N_0 \equiv  \frac{1}{[\frac{9}{4 \pi} \frac{M_p^2}{v^2} \xi]} \,\,\,  \mbox{  and  }  \,\,\, y \equiv \frac{\sqrt{V_0}}{ M_p v}  \,\,\, ,
\end{equation}
and temporarily work in units where $M_p=1$.   Working in the approximation $\xi \ll 1$, which will be accurate for the parameter choices that we
consider, we choose 
$N=60$, $n_s=0.9603$ and $\Delta_R = 4.69 \times 10^{-5}$.  Then, Eqs.~(\ref{eq:ns}), (\ref{eq:dsq}) and (\ref{eq:efold}) lead to the constraints:
\begin{equation}
0.9603 = 1+ \frac{1}{N_0} x_i
\label{eq:se1}
\end{equation}
\begin{equation}
4.69 \times 10^{-5} = \frac{2\sqrt{6}}{\sqrt{\pi}} \frac{ y N_0}{\sqrt{1-x_i^2}}
\label{eq:se2}
\end{equation}
\begin{equation}
60 = 2 N_0 \ln \left[\sqrt{\frac{(1-x_i)(1+x_f)}{(1+x_i)(1-x_f)}}\right] \,\,\, .
\label{eq:se3}
\end{equation}
The parameter $x_f$ is set by the scale $m_\chi$ and can be chosen freely, provided that the magnitude of the cosine is less 
than one.   For this example, we choose $x_f=0.8$.   Now the three equations above can be solved for the three unknowns, $N_0$, $x_i$ 
and $y$.   We find
\begin{align}
x_i &= -0.64 \,\, ,\nonumber \\
N_0 &= 16 \,\, ,\nonumber \\
y &= 8.0 \times 10^{-7} \,\, .
\end{align}
Once $v$ is specified, we can solve for the parameters $V_0$ and $m_0$ (the latter given by the definition of $\xi$.).  In this example, we choose 
$v=M_p/100$.  Then we find (including the input mass scales, for comparison)
\begin{align}
M_P &= 1.2 \times 10^{19} \mbox{ GeV}  \,\, ,\nonumber \\
v & = M_P/100 \,\, ,\nonumber \\
V_0 & = (1.1 \times 10^{15} \mbox{ GeV} )^4 \,\, , \nonumber \\
m_0 & = 6.7 \mbox{ TeV} \,\, .  \label{eq:scales1}
\end{align}
In our fundamental theory, Eq.~(\ref{eq:simpot}), $V_0$ is fixed by Eq.~(\ref{eq:c0}).  We find that the value for $V_0$ shown in 
Eq.~(\ref{eq:scales1}) is obtained for the dimensionless parameter choices\footnote{Given 
our normalization of the quartic couplings, perturbativity requires that they be $\ll (4\pi)^2/4! \approx 6.6$, which is easily satisfied.} 
$\lambda=0.1$, $\lambda_{12}=0.2$ and $c_1=0.051$.  Notice that none of the fundamental dimensionless couplings 
is forced to be unnaturally small, unlike the non-supersymmetric model based on the group $D_4$ that appeared in Ref.~\cite{Ross:2009hg};
the $D_4$ symmetry in that proposal allows marginal SO(2)-violating quartic self-couplings for the inflaton doublet, which necessitates a fine-tuning, 
while the $S_3$ symmetry prevents such operators and avoids this outcome. Given our choice of $x_f$, it follows from 
Eq.~(\ref{eq:ends}) that $m_\chi = 2.5 \times 10^{15}$~GeV.  Since this is a non-supersymmetric model, tuning of scalar masses is unavoidable; 
however, the $\chi$ mass is at a relatively high scale, so the largest  tuning required is still that of the Higgs boson mass, as in the 
standard model.    

Now we can summarize the values of the remaining cosmological parameters:
\begin{align}
\epsilon & =   7.9 \times 10^{-8} \,\,\, ,\nonumber \\
r      & = 1.3 \times 10^{-6} \,\,\, , \nonumber \\
n_r  & = 1.1 \times 10^{-3} \,\,\, .
\end{align}
These are consistent with the current bounds, assuming that one conservatively accepts the Planck upper bound on $r$.  An observable 
primordial gravitational wave signal, if confirmed, would rule out this parameter choice.   Therefore, we next consider how one could obtain a solution with 
larger $r$.

\section{Enhancing Primordial Gravity Waves} \label{sec:gw}

In the slow-roll approximation, by Eqs.~(\ref{eq:srp}) and (\ref{eq:r}), 
\begin{equation}
r=16 \, \epsilon=\frac{M_P^2}{\pi}\left(\frac{V'}{V}\right)^2 \,\,\, . 
\end{equation}
On the other hand, the scalar spectral index was given in Eq.~(\ref{eq:ns1}),
\begin{equation}
n_s(k)=1-6\epsilon+2\eta, \end{equation}
with value $n_s=0.9603\pm0.0073$, from Ref.~\cite{Ade:2013uln}. In order to increase $r$ with fixed $n_s$ in our model, we need to increase the 
values of both $\epsilon$ and $\eta$ at the time that the fluctuations were created, which we take to be $60$ $e$-folds prior to the end of inflation.  We 
must therefore increase $|V'/V|$ while $V''/V$ becomes less negative; this suggests that the inflaton in our model should 
minimize $|\cos(3\theta_i/v)|$ in order to obtain large $r$.  Although we find that it is challenging to obtain 60 $e$-folds of inflation while satisfying observational constraints beginning 
with such small magnitude of $\cos(3\theta_i/v)$, we find nonetheless that there are points in parameter space where a primordial gravitational 
wave signal is large enough to be potentially observable in upcoming experiments.  These points require a relatively small separation between $v$ and $M_P$, pushing the limits of effective field theory.    

\begin{figure}[h]
\includegraphics[width=2.5in]{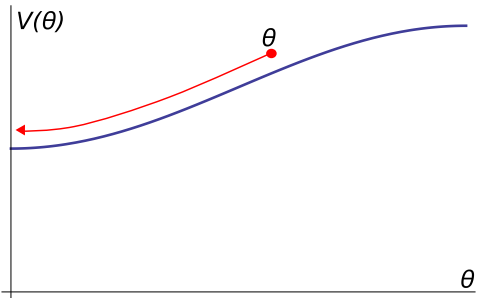}
\centering
\caption{The Lyth bound is evaded if the inflaton slowly rolls from a steep point in the  potential to near the minimum before the waterfall fields turn on.}
\label{fig:Vinf}
\end{figure}

The Lyth bound~\cite{Lyth:1996im} relates the number of $e$-folds of inflation to the change in the inflaton field $\theta$ during the same period, and suggests 
that in a wide class of models it is not possible to obtain a sizable gravitational wave signal without a change in the inflaton field during inflation that is much
larger than $M_P$. Such large field values would be  problematic for the effective-field-theory interpretation of  the model. Using the inflaton 
equation of motion and the relations for the power spectra of scalar and tensor modes in the slow-roll approximation, one obtains the relation~\cite{Lyth:1996im}, 
\begin{equation}
\left(\frac{d\theta}{dN}\right)^2=\frac{M_P^2}{64\pi}\,r \,\,\, .
\end{equation}
If $r$ is roughly constant during the last 60 $e$-folds of inflation, then one obtains,
 \begin{equation}
 \Delta\theta=\frac{1}{8\sqrt{\pi}} N\sqrt{r}\, M_P  \,\,\, ,
 \label{eq:Lyth}
 \end{equation}
which exceeds $M_P$ for $N \sqrt{r} > 8\sqrt{\pi}$.   In particular, this will be the case if $N=60$ and $r$ is of a typical observable value, for example $r \sim 0.1$.  We refer to Eq.~(\ref{eq:Lyth}) as the Lyth bound.
Hybrid natural inflation models, including the one presented here, can  evade the Lyth bound if the inflaton rolls from a steep point in the potential to 
near the bottom of the potential prior to the end of inflation~\cite{Hebecker:2013zda,Carrillo-Gonzalez:2014tia}, as sketched in Fig.~\ref{fig:Vinf}.  In that case $r$ 
varies significantly during inflation, which violates the assumption of nearly constant $r$ that fed into the bound. 

In order to obtain 60 $e$-folds of inflation in this enhanced-gravity-wave scenario, we need inflation to end near the bottom of the inflaton potential, which is possible if the waterfall fields 
have large diagonal masses $m_\chi$.  After fixing the parameters to the well-measured values of $\Delta_R$ and $n_s$, we find that the less-well-measured running of the scalar tilt, $n_r=-0.013\pm0.009$ from the Planck experiment \cite{Ade:2013uln}, in fact provides the greatest obstacle to rolling from near the steepest point of the potential.  A viable parameter choice within $2\sigma$ of the measured $n_r$ is obtained by setting $x_f = 0.995$ and $v=M_P/2$, in which case we 
find\footnote{For this point in parameter space, there is a more substantial difference in the second significant digit between the exact results and those obtained using the small-$\xi$
approximations in Eqs.~(\ref{eq:se1}), (\ref{eq:se2}) and (\ref{eq:se3}).  Hence, we show the exact results in this section.}
\begin{align}
x_i & = -0.32 \,\,\, , \nonumber \\
N_0 & = 9.3\,\,\, , \nonumber \\
y & = 1.70  \times 10^{-6}  \,\,\, .
\end{align}
The physical mass scales in this case are given by
\begin{align}
M_P & = 1.2 \times 10^{19} \mbox{ GeV}  \,\,\, , \nonumber \\
v & = M_P/2 \,\,\, ,\nonumber \\
V_0 & = (1.12 \times 10^{16} \mbox{ GeV})^4 \,\,\, ,\nonumber \\
m_0 &= 2.6 \times 10^6 \mbox{ GeV}  \,\,\, .
\end{align}
In terms of the fundamental potential, Eq.~(\ref{eq:simpot}), the scale $V_0$ can be reproduced in this case
with the choices $\lambda= 0.1$, $\lambda_{12}=0.2$ and  $c_1=0.0017$.  In that case, from Eq.~(\ref{eq:ends}) we find $m_\chi=1.8\times10^{17}$ GeV.  

The cosmological parameters
evaluated at $\theta=\theta_i$ are now
\begin{eqnarray}
&& \epsilon = 8.9 \times 10^{-4} \,\,\, ,\nonumber \\
&& r = 0.014  \,\,\, ,\nonumber \\
&& n_r = 4.8\times10^{-3} \,\,\, .
\end{eqnarray}
For this point in parameter space, a primordial gravitational wave signal could be within the reach of future CMB polarization measurements.   With the same 
value of $x_f$ but with $v=M_P/3$ rather than $M_P/2$, $r$ decreases to $0.0066$.  We have assumed that the cutoff of the theory is $M_P$, where quantum 
gravity effects are expected to become strong, rather than the reduced Planck mass $M_*=M_P/\sqrt{8\pi}$ that normalizes the gravitational coupling.  If we 
assume $v=M_*/2$ with the same value of $x_f$ as above, we obtain $r=0.00061$.  For comparison, the upcoming PIPER experiment expects a sensitivity to measure $r$ as low as 0.007
\cite{Lazear:2014bga}.

\section{Inflation Shut-off and Reheating} \label{sec:offreheat}

In this section, we consider the end of inflation and reheating.  We first present estimates that indicate the end of inflation happens abruptly\footnote{For alternatives
to this requirement, see Ref.~\cite{alt}.},  less than a Hubble time after the $\chi$ fields develop vacuum expectation values.

Our estimates follow the arguments of Ref.~\cite{Linde:1993cn}.  Consider the evolution of the inflaton field $\theta$ during 
$\Delta t = H^{-1}$ after the critical time $t_c$, where Eq.~(\ref{eq:ends}) is an exact equality.  At the very end of slow roll,
$3 H \dot{\theta} \approx -V^\prime(\theta_f)$; hence the change in the inflaton field during the subsequent $\Delta t$ is given by
\begin{equation}
\Delta\theta = -\frac{3  M_P^2}{8 \pi v} \frac{\xi \sin(3 \theta_f/v)}{[1-\xi \cos(3 \theta_f /v)]} \,\, .
\end{equation}
At $t_c$, the $\chi$ mass matrix has a zero eigenvalue, so the magnitude of the negative mass squared term that emerges $\Delta t$ later
is determined by $\Delta\theta$.  To assure a rapid evolution of the $\chi$ fields, we require that the magnitude of this 
negative squared mass is larger than $H^2$,
\begin{equation}
3 |c_1| \frac{v^2}{M_P} \sin(3 \theta_f/v) |\Delta \theta| > H^2  \,\, ,
\end{equation}
which, in the notation of the previous section, leads to the inequality 
\begin{equation}
\frac{27}{64 \pi^2} |c_1|\, \xi \,(1-x_f^2) \frac{v}{M_P} > \frac{V_0}{M_P^4}  \,\,\, .
\label{eq:ineq1}
\end{equation}
(Here and below we work to lowest order in $\xi\ll 1$.)  In addition, the non-zero $\chi$ vevs after $t_c$ generate a contribution 
to the $\theta$ mass squared which we also require to be greater than $H^2$,
\begin{equation}
9 |c_1| \frac{v}{M_P} \langle \chi_1 \chi_2 \rangle > H^2 \,\,\, ,
\end{equation}
which reduces to
\begin{equation}
\frac{27}{16 \pi} \, \frac{c_1^2 }{2\lambda+\lambda_{12}} (1-x_f) \,\left(\frac{v}{M_P}\right)^4 > \frac{V_0}{M_P^4} \,\,\,.
\label{eq:ineq2}
\end{equation}
For the two points in parameter space studied in Secs.~\ref{sec:num} and \ref{sec:gw}, respectively, we find numerically that the inequalities in
Eqs.~(\ref{eq:ineq1}) and (\ref{eq:ineq2}) are satisfied by between four and six orders of magnitude.  This suggests that the fields will
be driven to their global minimum sufficiently quickly, bringing inflation to an end.

The reheat temperature is sensitive to whether there is substantial preheating and depends on details of the couplings of the waterfall fields to matter, but for an estimate we assume reheating through a Higgs portal due to the quartic coupling, 
\begin{equation}
V_{\chi^2 H^2}=\frac{\lambda_{\chi H}}{2} (\chi_1^2+\chi_2^2)H^\dagger H  \supset\frac{\lambda_{\chi H}}{2} \chi^2 H^\dagger H \,\,\, ,
\end{equation}
where the waterfall field $\chi\equiv\left(\chi_1-\chi_2\right)/\sqrt{2}$ oscillates during reheating about the minimum of $V(\theta,\chi_i)$, which was determined in Eq.~(\ref{eq:chi2}).  We neglect the mixing with the orthogonal combination of $(\chi_1+\chi_2)/\sqrt{2}$ and the inflaton field $\theta$ in this simplified analysis.  The Higgs-portal coupling  contains the term $\lambda_{\chi H} \langle \chi \rangle \chi H^\dagger H$, where
$\langle \chi \rangle/\sqrt{2} = \langle\chi_1\rangle =-\langle \chi_2\rangle$, with the $\langle \chi_i \rangle$ determined by Eq.~(\ref{eq:chi2}). This coupling leads to the $\chi$ decay rate,
\begin{equation}
\Gamma_\chi=\frac{\lambda_{\chi H}^2\langle\chi_1\rangle^2}{4\pi m_{\chi\mbox{eff}}}.
\end{equation}
The effective $\chi$ mass at the minimum of the potential is given by, \begin{equation}
m_{\chi\mbox{eff}}^{2}=\frac{2 c_1 v^3}{M_P}-2m_\chi^2.
\end{equation}
In most scenarios the reheat temperature is  within an order of magnitude of \cite{Kofman:1997yn}\begin{equation}
T_{rh}\sim\sqrt{M_P\Gamma_\chi}=\frac{\lambda_{\chi H}}{4}\sqrt{\frac{M_P\,m_{\chi\mbox{eff}}}{(2\lambda+\lambda_{12})\pi}}.
\end{equation}
With parameters as in Sec.~\ref{sec:num} and Sec.~\ref{sec:gw} we find a generically high reheat temperature\footnote{Note that the coupling $\lambda_{\chi H}$
first affects the flatness of the inflaton effective potential at two-loops, but only if a Planck-suppressed inflaton-Higgs coupling is present.  Such a coupling can be taken 
small independently so that the range of $\lambda_{\chi H}$ is not restricted from this consideration.  All other effects on the inflaton potential involving $\lambda_{\chi H}$ 
occur at three or more loops.} $T_{rh}\sim 10^{17}\lambda_{\chi H}$ GeV. 

\section{Conclusions} \label{sec:conc}

We have analyzed a model of inflation based on the non-Abelian discrete group $S_3$.   The mass term and quartic self-coupling of a doublet of scalar fields 
preserve an accidental SO(2) symmetry. The SO(2)  is spontaneously broken, giving rise to a pseudo-Goldstone boson which plays the role of the inflaton, 
as in natural inflation.  After the inflaton rolls sufficiently, the coupling of the inflaton to two additional scalar fields generates an instability in a linear 
combination of those fields, ending inflation and reheating the universe as in hybrid inflation.  We studied constraints on the model due to the slow-roll 
conditions, the requirement of at least 60 $e$-folds of inflation, the measured magnitude of cosmic density perturbations, the measured scalar spectral 
index and its running. The model has a  viable parameter space with technically natural couplings, and can 
accommodate potentially observable power in tensor modes without super-Planckian field values during inflation, with $r\sim0.01$.  

Our work has been motivated in part by the minimality of $S_3$, which is the smallest possible non-Abelian discrete gauge group.  However, it also is worth pointing out that the 
group $S_3$ has been used successfully in flavor model building~\cite{s3papers}.  Such models include substantial scalar sectors (the flavons) that are restricted by the 
discrete symmetry.  It would be interesting in future work to see if the model described here could be incorporated into the flavor-symmetry-breaking sector of a flavor model 
involving $S_3$ symmetry.   In addition, the present model was constructed in a non-supersymmetric framework, for the sake of simplicity.   A study of a supersymmetric $S_3$ 
natural hybrid inflation model, which would also stabilize the electroweak scale, will be discussed elsewhere~\cite{ray}.

%%%%%%%%%%%%%%%%%%%%%%%%%%%%%%%%%%%%%%%%%%%%%%%%%%%%%%%%%%%
\begin{acknowledgments}  
C.D.C.,  J.E. and R.R. were supported by the NSF under Grant PHY-1068008.   The opinions and conclusions expressed herein 
are those of the authors, and do not represent the National Science Foundation.
\end{acknowledgments}
%%%%%%%%%%%%%%%%%%%%%%%%%%%%%%%%%%%%%%%%%%%%%%%%%%%%%%%%%%%     

%\appendix
%\section{}

\end{document}